\documentclass[11pt]{elsarticle}
\makeatletter
\def\ps@pprintTitle{%
 \let\@oddhead\@empty
 \let\@evenhead\@empty
 \def\@oddfoot{\centerline{\thepage}}%
 \let\@evenfoot\@oddfoot}
\makeatother

\usepackage{tikz,pgfplots}
\newif\ifExtended
\Extendedfalse
\pgfplotsset{compat=1.10}
\DeclareMathAlphabet{\mathpzc}{OT1}{pzc}{m}{it}
\usepackage[utf8]{inputenc}
\usepackage{verbatim,subcaption}
\usepackage[textsize=scriptsize]{todonotes}
\usepackage{colortbl}
\definecolor{Gray}{gray}{0.90}
\usepackage{pgf}
\usepackage{tikz}
\usepackage[framemethod=TikZ]{mdframed}
\usepackage{amsmath}
\usepackage{amssymb}
\usepackage{hyperref}

  \definecolor{dsblue}{RGB}{65,105,225}
  \definecolor{spgreen}{RGB}{46,139,87}
  \definecolor{tored}{RGB}{220,20,60}
  \definecolor{glyellow}{RGB}{218,165,32}
  \definecolor{botcolor}{RGB}{119,136,153}
  \definecolor{linecolor}{RGB}{128,128,128}
  \definecolor{topcolor}{RGB}{0,0,139}
  \tikzstyle{mypoint}=[shape=circle,draw,white,bottom color=botcolor,top color= topcolor]
  \tikzstyle{myline}=[linecolor,thin]
  \tikzstyle{mythick}=[thick]

\newcommand{\levels}{\mathcal{L}}

\newcommand{\targets}{\mathcal{T}}

\newcommand{\lm}{\lambda}

\newcommand{\DefMixed}{\Phi}

\newcommand{\pAw}{$\mathcal{A}$}

\newcommand{\AttackerMixed}{\Theta}

\begin{document}
\begin{frontmatter}

\title{Uncertainty in Cyber Security Investments}

\author[icl]{Andrew Fielder}
\author[ait]{Sandra K\"{o}nig}
\author[surrey]{Emmanouil Panaousis}
\author[ait]{Stefan Schauer}
\author[ait]{Stefan Rass}

\address[icl]{Imperial College London} 
\address[ait]{Austrian Institute of Technology}
\address[surrey]{University of Surrey}
\address[ukl]{Universit\"{a}t Klagenfurt}

\begin{abstract}
When undertaking cyber security risk assessments, we must assign numeric values to metrics to compute the final expected loss that represents the risk that an organization is exposed to due to cyber threats. 
Even if risk assessment is motivated from real-world observations and data, there is always a high chance of assigning inaccurate values due to different uncertainties involved (e.g., evolving threat landscape, human errors) and the natural difficulty of quantifying risk per se. 
Our previous work \cite{Fielder2016} has proposed a model and a software tool that empowers organizations to compute optimal cyber security strategies given their financial constraints, i.e., available cyber security budget. 
We have also introduced a general game-theoretic model \cite{Rass2015} with uncertain payoffs (probability-distribution-valued payoffs) showing that such uncertainty can be incorporated in the game-theoretic model by allowing payoffs to be random.
In this paper, we combine our aforesaid works and we conclude that although uncertainties in cyber security risk assessment lead, on average, to different cyber security strategies, they do not play significant role into the final expected loss of the organization when using our model and methodology to derive this strategies. 
We show that our tool is capable of providing effective decision support.
To the best of our knowledge this is the first paper that investigates how uncertainties on various parameters affect cyber security investments. 
\end{abstract}
\end{frontmatter}

%
%
\section{Introduction}
\label{sec:intro}
Many organizations do not have a solid foundation for an effective information security risk management.~As a result,~the increasingly evolving threat landscape in combination with the lack of appropriate cyber security defenses poses several and important risks. On the other hand, the implementation of an optimal cyber security strategy (i.e., formal information security processes; technical mechanisms; and organizational measures) is not a straightforward process.~In particular Small and Medium Enterprises (SMEs) are a priority focus sector for governments' economic policy. Given that the majority of SMEs are restricted by limited budgets for investing in cyber security,~the situation becomes cumbersome,~as without cyber security mechanisms in place,~they may be significantly impacted by inadvertent attacks on their information systems and networks leading,~in most cases,~to undesirable business effects.

Yet,~it is not only the limited budgets.~Even if these are available to some extent,~investing in cyber security is challenging due to the evolving nature of cyber threats that introduces serious uncertainties when undertaking cyber security risk assessments.~This asymmetry can highlight an investment decision from optimal to inefficient due to: (i) exploitation of newly found vulnerabilities that were not patched by the latest investment; and/or (ii) the mistaken values to risk assessment parameters,~which lead to erroneous optimal cyber security strategies.~The purpose of this paper is exactly that; ``to investigate how uncertainties in conducting cyber security risk assessment affect cyber security investments''. 

\subsection{Cyber security investments} 
According to a 2017 IBM report \cite{ibm2017report}, despite the decline (10\% percent) in the overall cost of a data breach over previous years to \$3.62 million, companies in this year's study are having larger breaches. 
A study conducted by the Ponemon Institute \cite{ponemon2015cost}, in 2015, on behalf of the security firm Damballa shows that although businesses spend an average of \$1.27 million annually and 395 people-hours each week responding to false alerts,~thanks to faulty intelligence and alerts,~breaches have actually gone up dramatically in the past three years.


The main challenges faced by organizations when it comes to investing in cyber security can be summarized as follows:

\begin{itemize}
\item lack of methods of determining accurate values for risk assessment parameters;
\item complexity of developing a holistic methodology that models an organization's environments,~performs risks assessment and finally derives an optimal investment solution; and
\item new threats emerge changing the level of risk derived prior to their appearance and therefore making the most recent investment non-optimal.
\end{itemize}

The literature of economics of security is quite rich and it comes to methodologies for investing in cyber security \cite{Lee2011,chronopoulos2017options,benaroch2017real,gordon2015increasing,moore2016identifying}. In our previous works \cite{Fielder2016}, \cite{Panaousis2014}~we compared different decision support methodologies for security managers to tackle the challenge of investing in security for SMEs.~To undertake the risk assessment of the proposed model,~we used \emph{fixed values} for the payoffs of the players (i.e.,~defender and attacker).~These values were set by using a mapping from the SANS Critical Security Controls \cite{web:SANS} combined with the Common Weakness Enumeration (CWE) Top 25 Software Vulnerabilities \cite{web:CWE}.~The data for this paper was published here \cite{Casestudy2015}.~Although the use of data from well-known sources made our risk assessment valid and important,~this approach ignored the fact that in real-world scenarios there is a very high amount of uncertainty when setting the payoff values.~And in fact,~even the data used in \cite{Fielder2016},~is just as accurate as the activities taken by experts when defining these values.~But such activities are prone to error due to: (i) being subjective to the human experience each time; (ii) the evolving threat landscape that unavoidably dictates new risk assessment values; and (iii) new assets being added to an organization's environment (i.e.,~infrastructure) therefore altering the current security posture of the organization.

\subsection{Decision under uncertainty} 
Decision problems often involve uncertainty about the consequences of the potential actions. Currently existing decision support methods use to either ignore this uncertainty or reduce existing information (e.g., by aggregating several values into a single number) to simplify the process. However, such approaches lose a lot of information. In \cite{Rass2015}, we introduce a game theoretic model where the consequences of actions and the payoffs are indeed random and, consequently, they are described as probability distributions. Even though the full space of probability distributions cannot be ordered, a subset of suitable \emph{loss distributions} that satisfy a few mild conditions can be totally ordered in a way that agrees with a general intuition of risk minimization. We show that existing algorithms from the case of scalar-valued payoffs can be adapted to the situation of distribution-valued payoffs. In particular, an adaption of the fictitious play algorithm allows computation of a Nash equilibrium for a zero-sum game. This equilibrium then represents the optimal way to decide among several options.
The model is described in more depth and illustrated with an example in \cite{Rass2016}. 

An area where such a framework is particularly useful is risk management. Risk is often assessed by experts and thus depends on many factors, including the risk appetite of the person doing the assessment. Additionally, the effects of actions are rarely deterministic but rather depend on external influences. Therefore, it is recommended by the German Federal Office of Information Security to do a \emph{qualitative} risk assessment which is consistent with our approach. We have applied the framework to model security risks in critical utilities such as a water distribution system in \cite{Busby2016}. In this situation, consequences are difficult to predict as consumers are not homogeneous and thus do not act like a single (reasonable) person. Another situation that can be modeled with this generalized game-theoretic approach is that of an advanced persistent threat (APT) \cite{Rass2017}. Recently, this type of attack has gained a lot of attention due to major incidents such as Stuxnet \cite{Karnouskos2011} or the attack on the Ukrainian power grid \cite{e-isac_analysis_2016}.

%
%
\section{Proposed Methodology}
Our work is inspired from two previous papers \cite{Fielder2016} and \cite{Rass2015} to investigate how uncertainties regarding cyber security risk assessment values affect the efficiency of cyber security investments that have been built upon game-theoretic and combinatorial optimization techniques (a multi-objective multiple choice Knapsack based strategy).~These uncertainties are reflected on the payoffs of the organization (henceforth refered to as the Defender). Although \cite{Fielder2016} was proven interesting and validated the UK's government aforesaid advice, it certainly did not account for uncertainties in the payoffs of the Defender. In real world scenarios,~defenders almost always operate with incomplete information,~and often a rough estimate on the relative magnitude of known cyber threats is the only information available to the cyber security managers.~Furthermore,~practical security engineers will argue that it is already difficult to obtain detailed information on risk assessment parameters. We envisage that by merging these two approaches,~we will be able to offer a decision support tool for cyber security investments with increased resiliency against threats facing SMEs.~More importantly,~our work addresses a wider class of cyber threats than \emph{commodity cyber threats},~which were investigated in \cite{Fielder2016}.~Although this assumption does not negate the possibility of zero-day vulnerabilities,~it removes the expectation that it is in the best interest of either player to invest heavily in order to either find a new vulnerability or be able to protect against these unknown vulnerabilities.~Therefore,~in the present paper,~we address even cyber attacks that target an organization with all means (i.e.,~advanced persistent threats).

\subsection{Cyber security Control Games with Uncertainty}
The Cyber security Control Games (CSCGs) developed so far \cite{Fielder2016} do not yet capture a problem that often arises in real life and especially in cyber security: a crisp prediction of the efficacy of cyber security controls as well as the values of the various other risk assessment parameters is often not possible.~Rather,~some intuitive information is available that describe some values as more likely than others.~In this paper,~we enrich the model recently presented in \cite{Fielder2016} by considering uncertainty in payoffs of the Defender (and of the Attacker since we play a zero sum game) in CSCG.~This is a two-stage cyber security investments model that supports security managers with decisions regarding the optimal allocation of their financial resources in presence of uncertainty regarding the different risk assessment values.

For a specific set of targets of the Attacker and security controls to be implemented by the Defender, our approach to cyber security risk assessment consists of two main steps. First, a zero-sum CSCG is solved to derive the optimal level at which the control should be implemented to minimize the expected damage if a target is attacked. This game accounts for uncertainty about the effectiveness of a control using probability-distribution as payoffs instead of crisp numbers. In previous work \cite{Rass2015}, we show that imposing some mild restrictions on these distributions admits the construction of a total ordering on a (useful) subset of probability distributions which allows to transfer solution concepts like the Nash equilibrium to this new setting. 

The most critical part in estimating the damage caused by a cyber security attack is predicting the efficacy of a control to protect a target $t$. Let us assume that we decide to implement the control at some level $l$; then we denote the efficacy of the control to protect target $t$ as $E(l,t)$.~Typically,~it is difficult to estimate this value,~even if $l$ and $t$ are known.~Thus,~we replace the exact value of $E(l,t)$ by a Gaussian distribution centered around the most likely value $e(l,t)$ with a fixed variance $\sigma^2$.~For simplicity, we assume that the uncertainty is equal for each cyber security control and implementation level.~This assumption can be relaxed if we have obtained an accurate value about the efficacy of a cyber security process (i.e., a control implemented at some level).~In order to avoid negative efficacy,~we truncate the Gaussian distributions to get a proper probability distribution on $[0,1)$. Allowing the efficacy of an implementation of a control at level $l$ on target $t$ to be random yields a random cyber security loss $S(l,t) = I(t)\,T(t)\,\left[1-E(l,t)\right]$.~This is the expected damage (e.g., losing some data asset) that the Defender suffers when $t$ is attacked and a control has implemented at level $l$.~This definition of loss is in line with the well-known formula,~risk = expected damage $I(t)$ $\times$ probability of occurrence $T(t)$ \cite{Oppliger2015}.~We assume that this loss will take values in a compact subset of $[1,\infty)$.~The losses in our games are thus random variables, so at this point, we explicitly deviate from the classical route of game theory. In particular, we \emph{do not} reduce the random payoffs to expected values or similar real-valued representatives. Instead, we will define our games to reward us in terms of a complete probability distribution, which is convenient for several reasons:

\begin{itemize}
  \item working with the entire probability distribution preserves all information available to the modeler when the games are defined. In other words, if empirical data or expertise on losses or utilities is available, then condensing it into a humble average sacrifices unnecessarily large amounts of information;
  \item it equips the modeler with the whole armory of statistics to define the payoff distribution, instead of forcing the modeler to restrict oneself to a ``representative value''. The latter is often a practical obstacle, since losses are not always easily quantifiable nor expressible on numeric scales (for example, if the game is about critical infrastructures and if human lives are at stake, a quantification in terms of ``payoff'' simply appears inappropriate).
\end{itemize}

Note that uncertainty in our case is essentially different to the kind of uncertainty that Bayesian or signaling games capture. While the latter is about uncertainty in the opponent the uncertainty in our case is about the payoff itself. The crucial difference is that Bayesian games nonetheless require a precise modeling of payoffs for all players of all types. This is only practically feasible for a finite number of types (though theoretically not limited to this). In contrast, our games embody an infinitude of different possible outcomes (types of opponents) in a single payoff, thus simplifying the structure of the game back into a standard matrix game, while offering an increased level of generality over Bayesian or signaling games.

In CSCG (a matrix game), Defender and Attacker have finite pure strategy spaces $\levels,\targets$ (where $l \in \levels, t \in \targets$) and a payoff structure of the Defender, denoted by $\mathbb{A}$, 
which in light of the uncertainties intrinsic to cyber security risk assessment, is a matrix of random variables.~During the game-play,~each player takes its actions at random,~which determines a row and column for the payoff distribution $F_{i,j}$. Repeating the game,~each round delivers a different random payoff $R_{ij}\sim F_{ij}$ whose distribution is conditional on the chosen scenario $i\in \levels,~j\in \targets$.~ Thus,~we obtain the function $F_{ij}(r)=\Pr(R_{ij}\leq r|i,j)$.~By playing mixed strategies,~the distribution of the overall expected random payoff $R$ is obtained from the law of total probability by
\begin{equation}\label{eqn:payoff-distribution}
    (F(\DefMixed,\AttackerMixed))(r) = \Pr(R\leq r) = \sum_{i,j}\Pr(R_{ij}\leq r|i,j)\cdot\Pr(i,j) = 
 \DefMixed^T\mathbb{A}\AttackerMixed,
 \end{equation}
when $\DefMixed,\AttackerMixed$ are the mixed strategies supported on $\levels,\targets$ and the player's moves are stochastically independent (e.g.,~no signaling).

Unlike classical repeated games,~where a mixed strategy is chosen to optimize a long-run average revenue,~equation \eqref{eqn:payoff-distribution} optimizes the distribution $F(\DefMixed,\AttackerMixed)$,~which is the same (identical) for every repetition of the game.~The game is in that sense static,~but (unlike its conventional counterpart) does not induce repetitions in practice,~since the payoffs are random (in each round),~but all having the same distribution.~Thus,~the ``distribution-valued payoff'' is always the same (whether there are repetitions of the game or not).

\subsection{Investment Optimization Problem with Uncertainty}
When having $c$ cyber security controls,~our plan for cyber investment is to solve $c$ CSCGs by splitting each of them up to a set of $m-1$ control subgames with $n$ targets and up to $\lm$ implementation levels for each control,~where $\lm \in \{1, \ldots, m\}$ (we set $\lm = 0$ to indicate that the control is not implemented at all).~For a CSCG the Control Subgame equilibria constitute the CSCG solution \cite{Fielder2016}.~Given the Control Subgame equilibria we then use a Knapsack algorithm to provide the general investment solution.~The equilibria provide us with information regarding the way in which each security control is \emph{best implemented},~so as to maximize the benefit of the control with regard to both the \pAw's strategy,~and the indirect costs of the organization.~For convenience,~we denote the Control Subgame solution by the maximum level of implementation available.~For instance,~for control $c_j$ the solution of Control Subgame $\mathcal{G}_{j\lm}$ is denoted by $Q^*_{j\lm}$.~Let us assume that for control $j$ the equilibria of all Control Subgames are given by the set $\{Q^*_{j0},\dots,Q^*_{jm}\}$.~For each control there exists a unique Control Subgame solution $Q_{j0}$,~which dictates that control $j$ should not be used.

We define an optimal solution to the Knapsack problem as 
$$\Psi = \{Q^*_{j\lm}\},~\forall\,j \in \{1,\dots,c\},~\forall\,\lm \in \{1,\dots,m\}$$.

\begin{figure}[h]
\centering
  \includegraphics[width=4.2in]{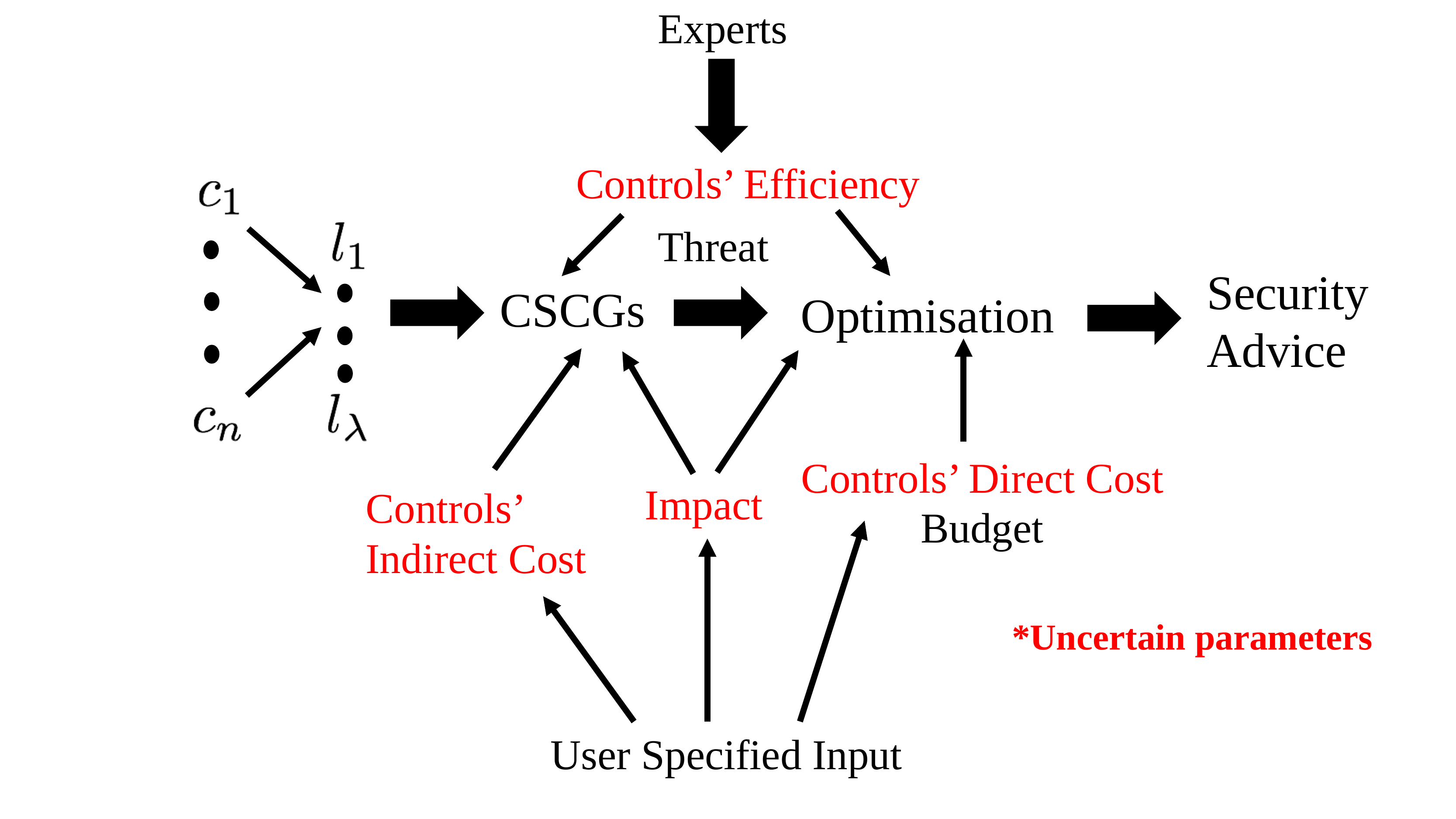}
  \caption{Overview of the cyber security investment methodology proposed in \cite{Fielder2016}.}
  \label{fig:investment_methodology}
\end{figure}

A solution $\Psi$ takes exactly one solution (i.e.,~equilibrium or cyber security plan) for each control as a policy for implementation.~To represent the cyber security investment problem,~we need to expand the definitions for both expected damage $S$ and effectiveness $E$ to incorporate the Control Subgame solutions.~Hence,~we expand $S$ such that $S(Q_{j\lm},t)$,~which is the expected damage on target $t$ given the implementation of $Q_{j\lm}$.~Likewise,~we expand the definition of the effectiveness of the implemented solution on a given target as $E(Q_{j\lm},~t)$.~Additionally,~we consider $\Gamma(Q_{j\lm})$ as the direct cost of implementing $Q_{j\lm}$.~If we represent the solution $\Psi$ by the bit-vector $\vec{z}$,~we can then represent the 0-1 Multiple Choice,~Multi-Objective Knapsack Problem as presented in (\ref{eq:investment}).

\begin{eqnarray}\label{eq:investment}
&&\max\limits_{\vec{z}}\frac{\sum_{i = 0}^n\Bigg\{\Big\{1 - \sum_{j=1}^{c}\sum_{\lm=0}^{m} E(Q_{j\lm},t_i)\,~z_{j\lm}\Big\}\,I(t_i)\,T(t_i)\Bigg\}}{t}\nonumber \\
&&\text{s.t.}~\sum_{j=1}^{c} \sum_{\lm=0}^{m}~\Gamma(Q_{j\lm}),z_{j\lm} \leq B \nonumber\\
&&~~~~\sum_{\lm=0}^{m} z_{j\lm} = 1,\,z_{j\lm} \in \{0,1\},\,\forall j = 1,\dots,c.
\end{eqnarray}
where $B$ is the available cyber security budget,~and $z_{j\lm}=1$ when $Q^*_{j\lm} \in \Psi$.~In addition,~we consider a tie-break condition in which if multiple solutions are viable,~in terms of maximizing the minimum,~according to the above function we will select the solution with the \emph{lowest cost}.~This ensures that an organization is not advised to spend more on security than would produce the same net effect. In Fig.~\ref{fig:investment_methodology}, we have illustrated the overview of the methodology followed to provide an optimal cyber security advice supporting decision makers with deciding about optimal cyber security investments. 

%
%
\section{Experiments}
\begin{figure}[h]
\centering
  \includegraphics[width=5.5in]{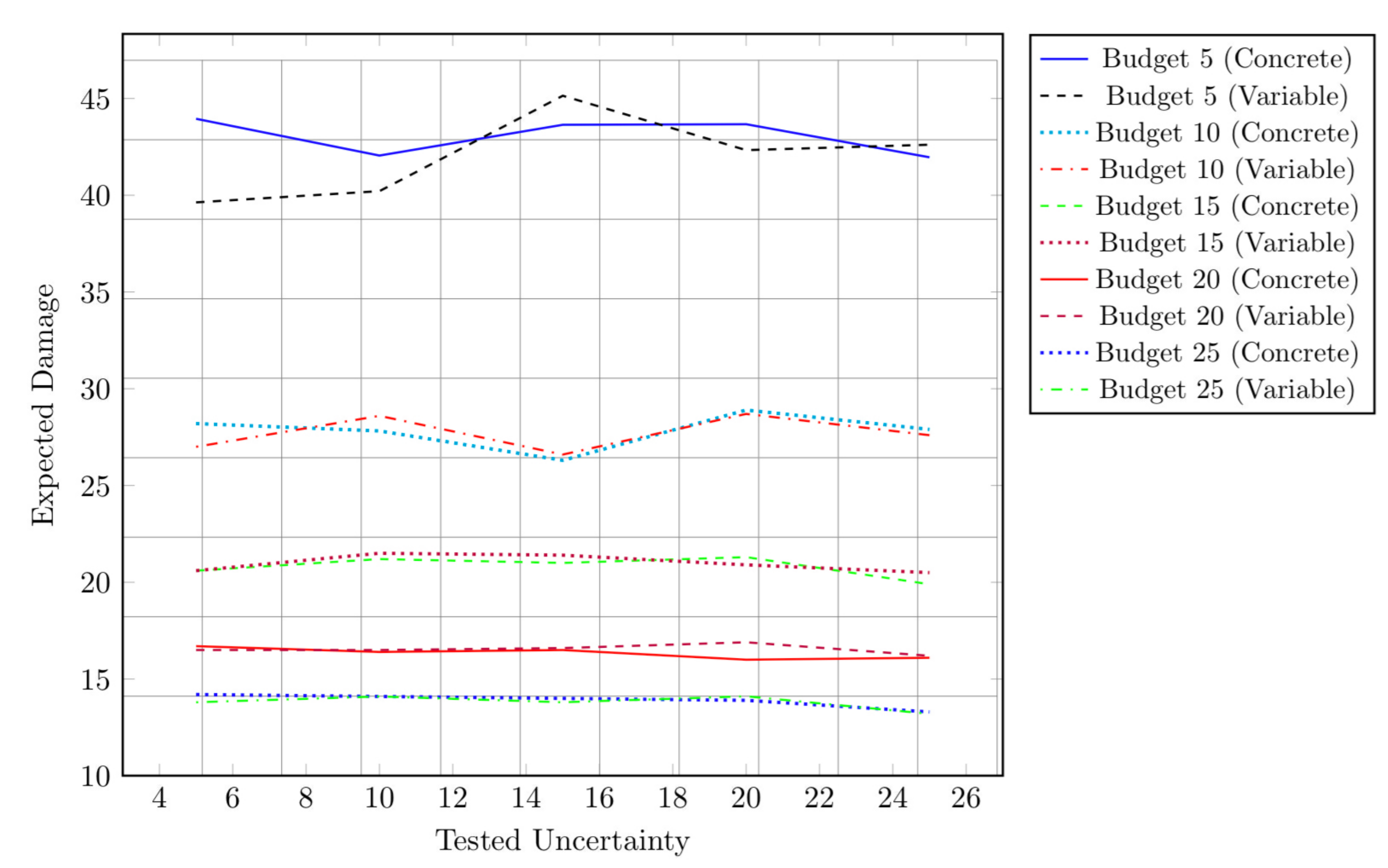}
  \caption{Expected damage tracked against uncertainty for each experimental configuration.}
  \label{fig_results}
\end{figure}

The results presented here represent the outcomes of experiments run using a test case comprised of a sample of 10 controls and 13 vulnerabilities from \cite{Casestudy2015} with different levels of uncertainty at each budget level.  All the reported results are collected in Fig. \ref{fig_results} and the expected damage is defined as a normalised value between 0 and 100.  In the following paragraphs, we will discuss the characteristics of each budget level.

The tables presented in this section present the best strategies seen at each budget level when tested with different levels of uncertainty.  The number represents the optimal level that a control should be implemented at, where 1 dictates the simplest possible configuration, 5 dictates the best but most restrictive possible configuration, and 0 represents no implementation of the control.

%
%
\textbf{Budget 5:}
The expected damage is distributed primarily between 35 and 45. 
With lower budgets, there are fewer viable solutions. 
There are few solutions that provide both good coverage and fall within the budget range, making the discovery of optimal solutions more difficult. The closer the direct cost of a solution tends towards the budget, the more likely the solution under uncertainty will exceed the budget and incur the penalty, this is prominent at the 5\% level of uncertainty.

With a very limited budget, the number of viable solutions are limited. With low uncertainty we see in Table \ref{tab_results_5}, all optimal solutions tend towards implementing only two controls.

With uncertainty greater than 0.2, we see a different solution, where the first control is implemented at a lower level, with the third control implemented at a higher level.

%
%
\textbf{Budget 10:} Unlike the lower budget level, we see that the average expected damage falls in the range of 26 to 29, which is half the range seen at budget 5. With more controls available, the expected damage should go down, however at the same time we see that the solutions become more consistent. The standard deviation is less than 2.5, with a difference in means that never exceeds 2.

Table \ref{tab_results_10} shows that the optimal results for budget 10 build on the basic pattern from those at budget 5, suggesting implementations for both controls 1 and 3 regardless of the level of uncertainty. With low uncertainty, control 9 is considered optimal, but at higher levels of uncertainty, controls 7 and 10 are considered optimal.

%
%
\textbf{Budget 15:} For a budget of 15, we see that the mean expected damage is between 19 and 22. At this budget and higher, we see that the difference in means between the certain and uncertain solutions never exceeds 1.
With the increased budget over the previous results, the optimal solution in Table \ref{tab_results_15}, now always considers a combination of the first three controls, where the rest of the budget is used to sporadically patch the worst remaining vulnerabilities as dictated by uncertainty. This means that at lower levels of uncertainty control 4 is preferred, while at higher levels of uncertainty, we see that control 10 becomes the favoured addition to the base set of controls, with control 9 preferred at 10\% uncertainty.

\begin{table}[!htb]
  \centering
    \caption{Optimal Solutions for Budget=5}
         \begin{tabular}{|l|llllllllll|l}
        \hline
            Uncertainty & 1 & 2 & 3 & 4 & 5 & 6 & 7 & 8 & 9 & 10 \\ \hline
            0\%          & 4 & 0 & 1 & 0 & 0 & 0 & 0 & 0 & 0 & 0 \\
            5\%          & 4 & 0 & 1 & 0 & 0 & 0 & 0 & 0 & 0 & 0 \\
            10\%        & 4 & 0 & 1 & 0 & 0 & 0 & 0 & 0 & 0 & 0 \\
            15\%        & 4 & 0 & 1 & 0 & 0 & 0 & 0 & 0 & 0 & 0 \\
            20\%        & 3 & 0 & 2 & 0 & 0 & 0 & 0 & 0 & 0 & 0 \\
            25\%        & 3 & 0 & 2 & 0 & 0 & 0 & 0 & 0 & 0 & 0 \\
            \hline
        \end{tabular}
        \label{tab_results_5}
\end{table}

\begin{table}[!htb]
  \centering
    \caption{Optimal Solutions for Budget=10}
         \begin{tabular}{|l|llllllllll|l}
        \hline
              Uncertainty & 1 & 2 & 3 & 4 & 5 & 6 & 7 & 8 & 9 & 10 \\ \hline
            0\%          & 3 & 0 & 1 & 0 & 0 & 0 & 0 & 0 & 4 & 0 \\
            5\%          & 3 & 0 & 1 & 0 & 0 & 0 & 0 & 0 & 4 & 0 \\
            10\%        & 3 & 0 & 1 & 0 & 0 & 0 & 0 & 0 & 4 & 0 \\
            15\%        & 4 & 0 & 3 & 0 & 0 & 0 & 1 & 0 & 0 & 0 \\
            20\%        & 4 & 0 & 3 & 0 & 0 & 0 & 1 & 0 & 0 & 0 \\
            25\%        & 4 & 0 & 2 & 0 & 0 & 0 & 0 & 0 & 0 & 2 \\            \hline
        \end{tabular}
        \label{tab_results_10}
\end{table}

\begin{table}[!htb]
       \centering
        \caption{Optimal Solutions for Budget=15}
        \begin{tabular}{|l|llllllllll|l}
            \hline
            Uncertainty & 1 & 2 & 3 & 4 & 5 & 6 & 7 & 8 & 9 & 10 \\ \hline
            0\%         & 4 & 2 & 3 & 1 & 0 & 0 & 0 & 0 & 0 & 0 \\
            5\%         & 4 & 2 & 3 & 1 & 0 & 0 & 0 & 0 & 0 & 0 \\
            10\%        & 4 & 2 & 3 & 0 & 0 & 0 & 0 & 0 & 1 & 0 \\
            15\%        & 4 & 3 & 2 & 0 & 0 & 0 & 0 & 0 & 0 & 1 \\
            20\%        & 4 & 3 & 2 & 0 & 0 & 0 & 0 & 0 & 0 & 1 \\
            25\%        & 4 & 3 & 3 & 0 & 0 & 0 & 0 & 0 & 0 & 1 \\
        \hline
    \end{tabular}
     \label{tab_results_15}
    \end{table}

     \begin{table}[!htb]
      \centering
        \caption{Budget=20}
        \begin{tabular}{|l|llllllllll|l}
        \hline
            Uncertainty  & 1 & 2 & 3 & 4 & 5 & 6 & 7 & 8 & 9 & 10 \\ \hline
            0\%          & 5 & 3 & 2 & 1 & 0 & 0 & 0 & 0 & 0 & 1 \\
            5\%          & 5 & 3 & 2 & 1 & 0 & 0 & 0 & 0 & 0 & 1 \\
            10\%         & 5 & 3 & 2 & 1 & 0 & 0 & 0 & 0 & 0 & 1 \\
            15\%         & 5 & 4 & 2 & 0 & 0 & 0 & 0 & 1 & 2 & 0 \\
            20\%         & 5 & 4 & 2 & 0 & 0 & 0 & 0 & 1 & 2 & 0 \\
            25\%         & 5 & 4 & 3 & 0 & 0 & 0 & 0 & 0 & 3 & 0 \\
        \hline
    \end{tabular}
    \label{tab_results_20}
    \end{table} 

%
%
\textbf{Budget 20:} The range of average expected damage is limited to less than 1, with the biggest discrepancy between certain and uncertain solution at the 20\% uncertainty level.

The optimal solutions from Table \ref{tab_results_20} add little to the general pattern of solutions that precede it, implementing the first 3 controls at varying levels. This is the only time that we see the optimal solution suggest the highest level of implementation for control 1. Here, control 10 is preferred at lower levels of uncertainty. At higher levels, this and control 4 are replace by a combination of controls 7 and 8.

\begin{table}[!htb]
    \caption{Solutions}
          \centering
        \caption{Optimal Solutions for Budget 25.}
        \begin{tabular}{|l|llllllllll|l}
            \hline
            Uncertainty & 1 & 2 & 3 & 4 & 5 & 6 & 7 & 8 & 9 & 10 \\ \hline
            0\%         & 4 & 4 & 3 & 3 & 0 & 0 & 0 & 3 & 0 & 0 \\
            5\%         & 4 & 4 & 3 & 3 & 0 & 0 & 0 & 3 & 0 & 0 \\
            10\%        & 4 & 4 & 3 & 3 & 0 & 0 & 0 & 3 & 0 & 0 \\
            15\%        & 4 & 4 & 1 & 3 & 0 & 0 & 0 & 2 & 3 & 0 \\
            20\%        & 4 & 4 & 1 & 3 & 0 & 0 & 0 & 2 & 3 & 0 \\
            25\%        & 3 & 2 & 3 & 3 & 0 & 0 & 1 & 0 & 4 & 0 \\
        \hline
    \end{tabular}
      \label{tab_results_25}
  \end{table}    
  
  \begin{table}[!htb]
    \caption{Solutions}
          \centering
        \caption{Base Solutions for All Budget Tested.}
        \begin{tabular}{|l|llllllllll|l}
            \hline
             Budget    & 1 & 2 & 3 & 4 & 5 & 6 & 7 & 8 & 9 & 10 \\
            5         & 3 & 0  & 1 & 0 & 0 & 0 & 0 & 0 & 0 & 0 \\
            10        & 3 & 0 & 1 & 0 & 0 & 0 & 0 & 0 & 0 & 0 \\
            15        & 4 & 2 & 2 & 0 & 0 & 0 & 0 & 0 & 0 & 0 \\
            20        & 5 & 3 & 2 & 0 & 0 & 0 & 0 & 0 & 0 & 0 \\
            25       & 3 & 2 & 1 & 4 & 0 & 0 & 0 & 0 & 0 & 0\\
            \hline
    \end{tabular}
        \label{tab_baseline}
  \end{table}

%
%
\textbf{Budget 25:} Considering the highest budget tested, we see that the average expected damage has a range of 1, between 13.2 and 14.2. This results in a difference in means of at most 0.4 and a minimum of 0.025. This is combined with standard deviations of no greater than 1.2 to provide consistent results between certain and uncertain solutions.

From Table \ref{tab_results_25}, the main difference in solutions is that control 4 becomes a permanent suggestion for implementation in addition to the other 3 core controls. Up to 20\% uncertainty, we see some variation of 6 controls, with consistent solutions up to 10\% uncertainty and a common solution at 15\% and 20\% uncertainty.

At 25\% uncertainty we see that the optimal solution deviates away from those solutions below. As with all of the results, despite a different solution, we still see a similar expected damage with the solution created in certain space. With uncertainty and a wide range of available configurations, it is reasonable to consider that there will be a number of solutions that offer similar results. Given that it still shares common factors, we can consider that most of the mitigation is handled by those four controls. The mitigation of the additional controls covers the change in values caused by uncertainty, this is similar to the case seen at 15\% uncertainty.

The following section highlights a number of common themes across the results, considering the expected results as well as themes consistent with the optimal solutions.

%
%
\section{Discussion}
Across all of the results in Fig. \ref{fig_results}, we see only a \emph{small difference in mean expected damage} between the optimal results with certain and uncertain parameters. This is represented by a difference in the mean values of comparable results not exceeding one standard deviation. While some of the consistency is due to multiple evaluations of solutions, the nature of the designs of the solutions similarly reduces the impact. The hybrid optimisation approach requires multiple different negative perturbations on values to be offset by positive perturbations on other controls before the impact will be seen. The value suggested by the expected damage captures these differences in the deviation of the results from the mean. 

The optimal results demonstrate a number of changes to the investment strategy as the uncertainty increases. This change can be explained as a combination of the factors that are uncertain. In general, this will be as a result of some controls becoming more effective than others at similar tasks. Less common results will have optimal solutions that might not be considered valid under a certain set of parameters, but based on uncertainty in the costs, would appear to be genuine.

It is with this last point that we find one of the sources for deviation in the average expected damage seen in the previous section. Above, we discuss having potentially invalid solutions seen to optimal, but we also need to consider the case, where the most optimal solution was eliminated due to potentially having a cost that would exceed the budget. 

Uncertainty in the cost is represented most prominently in the results at low budgets. This is due to the number of viable solutions that can be tested, since most solutions will exceed the budget. With this, the search space for solutions features more local optima, with less coherent strategies for traversal.

The consistency in the results can be explained by the coverage of certain controls and their effectiveness at completing that task. Across all the results displayed in Tables \ref{tab_results_5}, \ref{tab_results_10}, \ref{tab_results_15} and \ref{tab_results_20}, we see that control 1 is always selected, and with some limited exceptions, so is control 3. This gives us an impact on multiple vulnerabilities tested, causing a reduction in the expected damage. It is only at higher budgets, we see that the impact of multiple controls better filling the role of control 3 causes it to be replaced in the optimal solutions.

In addition to the idea that we see consistent results across low levels of uncertainty, we also see that the results identify that although there are a number of differences in the precise optimal solution, there is commonality among all of the optimal solutions present.

The trial was performed with a small set of attacks and controls. Increasing the number of controls and vulnerabilities could increase the potential for less consistent solutions, due to more overlap of controls. Regardless of the composition, good coverage of attack vectors is achieved as the optimal set of controls will always aim to mitigate the most expected damage across all targets.

A desired outcome of the experimental work was to see the extent of the commonality of optimal solutions for each of the levels of uncertainty. As has been explained above, we see that there are a number of commonalities, especially at the same budget levels. Table \ref{tab_baseline} shows the minimal set of controls and levels that are implemented regardless of the uncertainty.

In comparison to the optimal results for each of the budget levels, we see that these share common features on the first three controls, and later control 4. It is these controls that provide a base coverage of the attack vectors, as described previously. The worst performing base is that of budget 10, which reflects that of the budget 5, this is due to the deviation between low uncertainty and high uncertainty solutions.

From the cyber security perspective, we consider that there are sets of advice such as the UK's Cyber Essentials, that promote a number of controls. These pieces of advice suggest a set of controls that are reasonable to implement regardless of the degree of complexity or available budget. The base solutions shown here offer the same approach, demonstrating, what a solution should contain based on a constrained budget and uncertainty. These base solutions should be taken as a reference point for building secure systems, with decisions made regarding company specific requirements.

\section{Conclusions and future work}
This work extended previous work published in the field of decision support for cyber security. 
It has demonstrated an approach to cyber security investments under uncertainty, where a previous risk assessment based model was extended for this purpose. 
To explore this, a series of experiments looking at optimal cyber security investments under uncertainty were performed.
Uncertainty is naturally a challenge that all cyber security managers face when they have to take decisions. 
The derivation of exact values for various risk assessment parameters seems like an impossible task. 
Our work here highlights, that even with some uncertainty in factors that impact payoffs and viable strategies, there is consistency in the outcomes, where the majority of damage was being mitigated by only a few cyber security controls.
Although we have concluded to a set of numerical results that clearly demonstrate the benefit of our model and methodology, the expected extension of this work, would be to apply the proposed tools to a full realistic case study, allowing for a comparison to expert judgments, capturing where and how the uncertainty arises.

\bibliographystyle{IEEEtran}
\bibliography{refs}
\end{document}